\documentclass[fleqn,12pt,twoside,english]{article}
\usepackage{espcrc1}

\usepackage{graphicx}
\usepackage[figuresright]{rotating}
\usepackage{graphicx}
\usepackage{epsfig}
\usepackage{subfigure}
\usepackage{floatflt}
\usepackage{graphicx}
\usepackage{setspace}
\usepackage{babel}

\newcommand{\AmS}{{\protect\the\textfont2
  A\kern-.1667em\lower.5ex\hbox{M}\kern-.125emS}}

\title{Probing bulk properties and partonic collectivity via multi-strange baryons in Au+Au collisions at top RHIC energy.}

\author{Magali Estienne \address[MCSD]{{\small{SUBATECH, 4 rue Alfred Kastler, BP 20722, 44307 Nantes-cedex 3, FRANCE}}} (for the STAR Collaboration)}%

\begin{document}

\maketitle

\begin{abstract}
The study of the multi-strange baryon production in the final state of ultra-relativistic Au+Au collisions at $\sqrt{s_{NN}}=200GeV$ at RHIC gives information on the freeze-out conditions and probably on earlier stages of the collision. 
Chemical freeze-out temperature $T_{ch}$ as well as the strangeness phase space occupancy factor $\gamma_{s}$ extracted from statistical models are studied as a function of the collision centrality. $\gamma_{s}$ saturates at 1 in most central collisions suggesting an equilibration of strange quarks in the medium. A study of the collective motion of the collision is possible in the framework of a hydrodynamically-inspired model which considers the particles to be emitted from a locally thermalized system. In this view, multi-strange particles seem to develop significant radial transverse flow but smaller than $\pi$,$K$,$p$. Furthermore, $\Xi^{-}+\overline{\Xi}^{+}$ and $\Omega^{-}+\overline{\Omega}^{+}$ present an elliptic flow as strong as for previously measured baryons supporting the idea that a fraction of the final collective motion has been developed at an early partonic stage of the collision.
\end{abstract}

\section{INTRODUCTION}
The Lattice Quantum Chromodynamics calculations predict the existence of a transition between a hadronic gas 
and a quark gluon plasma at a temperature  around 170MeV in a domain close to the net-baryon free region \cite{Fodor03}. 
This deconfined state of quarks and gluons is expected to be formed in ultra-relativistic heavy ion collisions. The study of the final hadronic state properties of such
collisions essentially dominated by the low $p_{T}$ part of the particle spectra (the bulk)  via multi-strange particles may provide information on its dynamics from the early stage to the chemical and thermal freeze-out (FO). \\
Strange quarks whose mass is comparable to the temperature of the QGP formation are expected to be abundantly produced in the high temperature QGP phase because of parton rescattering and should achieve equilibration \cite{Raf03}. Hadronization process has been very well described in the framework of statistical model \cite{Raf03,PBM99,Bec04} by adjusting four free parameters, the chemical temperature $T_{ch}$, the baryon and strange chemical potential $\mu_{B}$ and $\mu_{s}$ and the strangeness phase space occupancy factor $\gamma_{s}$. This latter provides a measurement of the degree of strangeness equilibration in the system. Its evolution with centrality as well as $T_{ch}$ evolution gives a quantitative measurement of strangeness evolution in the bulk matter. \\
A study of the collective motion of the collision in the framework of a hydrodynamically-inspired model has previously shown that $\pi$, K, p particles seem to take part to the same transverse collective flow and freeze-out kinetically at a temperature smaller than $T_{ch}$ suggesting an expansion and the cooling of the system between chemical and thermal FO. Concerning multi-strange baryons, it has been suggested that these particles should not develop such significant transverse radial flow due to their presumably small cross section so that they should decouple much earlier in the collision \cite{nxu98,cheng03}. Their observed transverse radial flow would then primarily reflects partonic flow behaviour. Elliptic flow due to the initial asymmetry of the system in non-central collisions has also proven to be a good tool for understanding the properties of the early stage of the collisions \cite{Olli92}. Thus multi-strange baryon elliptic flow could be a valuable probe of the initial partonic system. As flow is an additive quantity, we present both radial transverse flow and elliptic flow measurements of multi-strange baryons in order to disentangle its hadronic and partonic contributions.

\section{RESULTS AND DISCUSSION}
\subsection{Bulk chemical properties}
\vspace*{2mm}
\begin{floatingfigure}[r]{8cm}%
\begin{spacing}{0.8}
\hspace*{-0.3in}
\vspace*{-0.3in}
\includegraphics*[ width=10cm,height=9cm, 
  keepaspectratio]{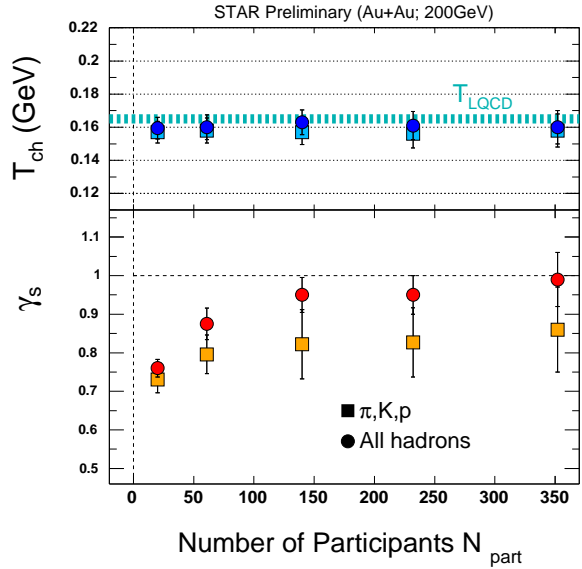}
\vspace*{-.15in}
\caption{Evolution of chemical temperature $T_{ch}$ (top) and strangeness phase space occupancy factor $\gamma_s$ (bottom) as a function of the number of participants. $T_{ch}$ and $\gamma_s$ have been calculated from statistical fits including $\pi$, $K$, $p$ (squares) and  $\pi$, $K$, $p$, $\Lambda$, $\Phi$, $\Xi$, $\Omega$ particles (circles).}
\label{fig:gammas}
\vspace*{.20in}
\end{spacing}
\end{floatingfigure}%

All the data presented in this contribution have been collected by the STAR detector described in \cite{Star03}. Multi-strange particles are identified via the topology of their decay $\Xi$$\rightarrow$$\Lambda+\pi$ and $\Omega$$\rightarrow$$\Lambda+K$ then the subsequent decay $\Lambda$$\rightarrow$$p+\pi$ with the respective branching ratios 100$\%$, 68$\%$ and 64$\%$. For more details see \cite{Effi02}. Corrections for tracking efficiency and detector acceptance were applied. The final corrected transverse momentum distributions are fitted in order to extract yields and inverse slope parameters.


The results of a statistical fit of particle ratios including $\pi^{\pm}$, $K^{\pm}$, $p$, $\overline{p}$, $\Lambda$, $\overline{\Lambda}$, $\Phi$, $\Xi^{\pm}$ and $\Omega^{\pm}$ are presented on Figure~\ref{fig:gammas}. Very good agreement between our data and the model are achieved for each centrality range at $\sqrt{s_{NN}}=200GeV$ at RHIC \cite{PBM03}. A fit has been performed first including $\pi^{\pm}$, $K^{\pm}$, $p$, $\overline{p}$. For the most central collisions, the four free parameters of the fit are $T_{ch}=157\pm6MeV$, $\mu_{B}=22\pm4MeV$, $\mu_{s}=3.8\pm2.6MeV$ and $\gamma_{s}=0.86\pm0.11$. The evolution of $T_{ch}$ and $\gamma_{s}$ with centrality are represented as square symbols on Figure~\ref{fig:gammas}. An other fit has also been performed including then all the hadrons $\pi^{\pm}$, $K^{\pm}$, $p$, $\overline{p}$, $\Lambda$, $\overline{\Lambda}$, $\Phi$, $\Xi^{\pm}$ and $\Omega^{\pm}$. The parameters obtained for the most central collision are $T_{ch}=160\pm5MeV$, $\mu_{B}=24\pm4MeV$, $\mu_{s}=1.4\pm1.6MeV$ and $\gamma_{s}=0.99\pm0.07$. $T_{ch}$ and $\gamma_{s}$ evolutions with centrality are represented as circles. We note no dependence of $T_{ch}$ with centrality. All the particles seem to chemically freeze-out at a temperature of 160$\pm$5MeV close to LQCD predictions, $T_{ch}$ seems to be essentially fixed by the most numerous $\pi$, $K$, $p$ particles and does not seem to be dependent on the initial system size. From peripheral to central collisions, including (multi-)strange hadrons in the fit, $\gamma_{s}$ increases from 0.8 and saturates at 1. This value suggests that in most central collisions at top RHIC energy, the phase space is saturated in strange quarks so that the system is close to strangeness equilibration. This increase of $\gamma_{s}$ also signs the existence of significant $s$$\overline{s}$ production processes at a partonic level such as gluon fusions. 
\vspace*{3mm}
\subsection{Collision dynamics}
\vspace*{2mm}

A hydrodynamically-inspired fit known as ``blastwave fit'' \cite{Blas93} assuming all particles are emitted from a thermal expanding source with a transverse flow velocity $<$$\beta_{T}$$>$ at the thermal freeze-out temperature $T_{fo}$ has been performed on $\pi$, $K$, $p$ spectra together and on $\Xi$ and $\Omega$ separately. A velocity profile $\beta_{T}(r)$=$\beta_{s}(r/R)^{n}$ was used, where $R$ is the radius of the source and $n$ was determined from the fit to the $\pi$, $K$ and $p$ spectra ranging from $n$=0.81 for the most central bin to $n$=1.42 for the most peripheral. For ($\pi$, $K$, $p$), 9 bins of centrality indexed from 1 (most central) to 9 (most peripheral) have been considered while 5 centrality bins for $\Xi^{-}+\overline{\Xi}^{+}$ have been studied. For $\Omega^{-}+\overline{\Omega}^{+}$, only the most central bin has been investigated. The results of the fits are presented on Figure~\ref{fig:blastwave}.
\begin{floatingfigure}[r]{9.cm}%
\begin{spacing}{0.8}
\hspace*{-0.2in}
\includegraphics*[ width=9.5cm,
  keepaspectratio]{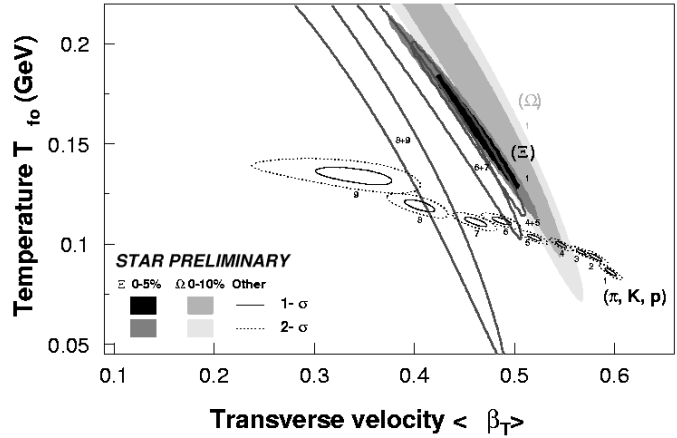}
\vspace*{-.15in}
\hspace*{.5in}
\caption{Kinetic freeze-out temperature, $T_{fo}$, as a function of the transverse flow velocity $<\beta_{T}>$ extracted from a hydro-inspired model of blastwave type from transverse momentum distributions.}
\label{fig:blastwave}
\vspace*{.20in}
\end{spacing}
\end{floatingfigure}%
The one and two sigma contours are represented for the best fit values ($T_{fo}$,$<$$\beta_{T}$$>$). For the most central bin, we note : 1) there is no overlap of the contours for ($\pi$,$K$,$p$) and ($\Xi$) suggesting that ($\pi$,K,p) take part to the same collective transverse radial flow different from the one developed by the $\Xi$ ; 2) concerning $\Xi$, they seem to thermally freeze-out at a temperature of $T_{fo}\sim153MeV$, close to the chemical FO temperature previously obtained from statistical fits to particle ratios whereas $T_{fo}$ for ($\pi$,$K$,$p$) amounts $90MeV$. It suggests that multi-strange particles should have decoupled earlier in the collision close to chemical FO ; 3) furthermore, the fact that they develop an as significant flow as $\Lambda$ and that their interaction cross-section is presumably very small, suggests that their flow has been developed prior to chemical FO so prior to the hadronization, probably at a partonic stage of the system and not in the hadronic phase as for ($\pi$,$K$,$p$). Otherwise, it is corroborated by the fact that the thermal FO parameters of the multi-strange baryons do not depend on the centrality and that $T_{ch}$ is close to $T_{fo}$. Concerning ($\pi$,$K$,$p$), results show that $T_{ch}$$>$$T_{fo}$ and that the difference between these two temperatures increases with centrality. It suggests a longer duration time between these two FO for the lightest particles ($\pi$,$K$,$p$) essentially due to their rescattering in the hadron phase while the system is cooling down.
These results indicate that Au+Au collisions with different initial conditions evolve always to the same chemical FO temperature, and then cool down further to a kinetic FO dependent on centrality.

\begin{figure}[htb]
\begin{minipage}[t]{80mm}
{\rule[10mm]{-2mm}{52mm}\epsfig{figure=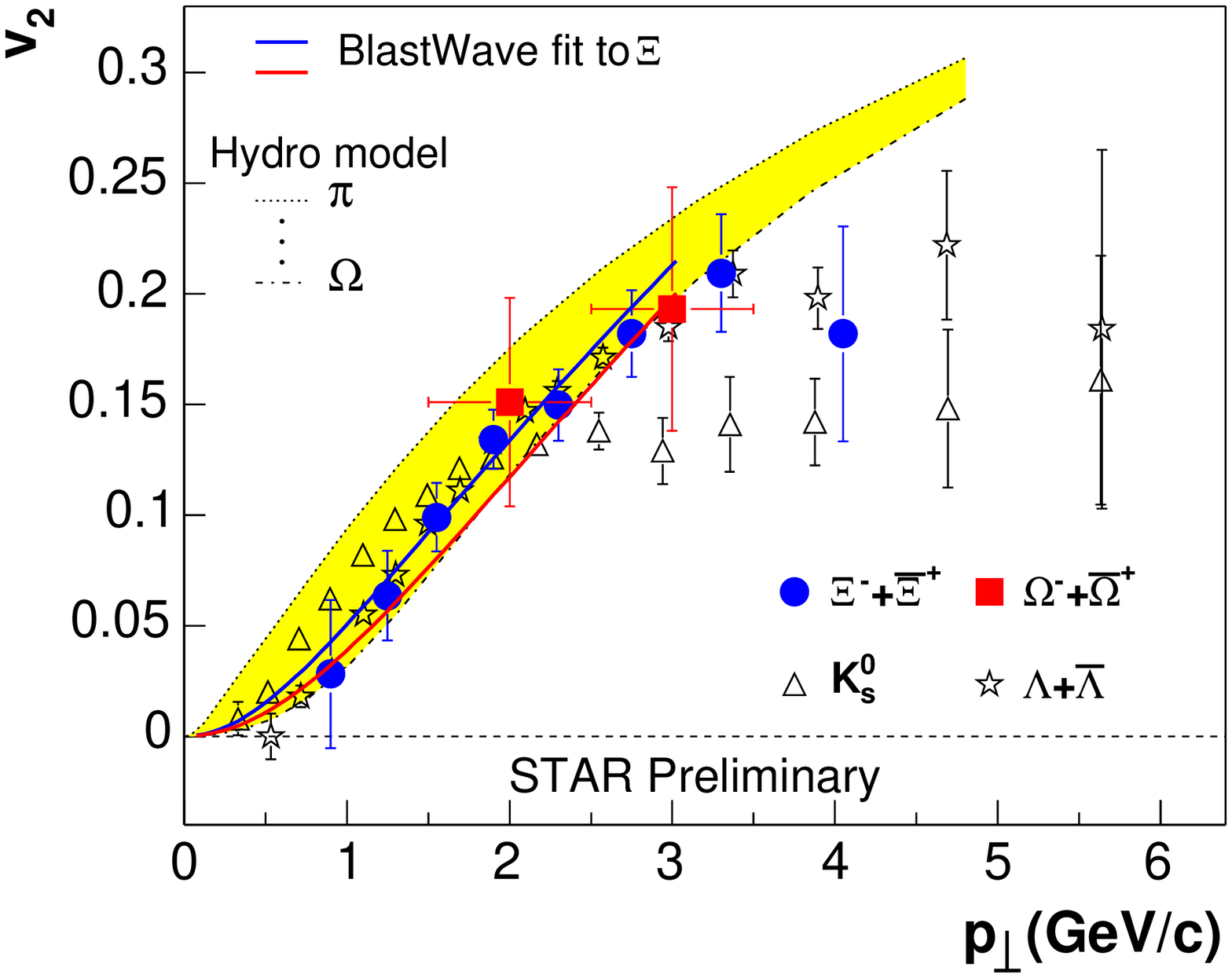,height=6cm}}
\hspace{-1cm}
\vspace*{-10mm}
\caption{Elliptic flow $v_{2}$ of $K_{s}^{0}$, $\Lambda+\bar{\Lambda}$, $\Xi^{-}+\bar{\Xi}^{+}$ and $\Omega^{-}+\bar{\Omega}^{+}$ from 200GeV Au+Au minimum bias collisions. Hydrodynamic model calculations are shown (colored zone) as well as hydro-inspired model calculations (thick lines) but are not commented in this letter \cite{QM04}.}
\label{fig:v2Pt}
\vspace*{-6mm}
\end{minipage}
\hspace{\fill}
\begin{minipage}[t]{75mm}
{\rule[10mm]{-2mm}{52mm}\epsfig{figure=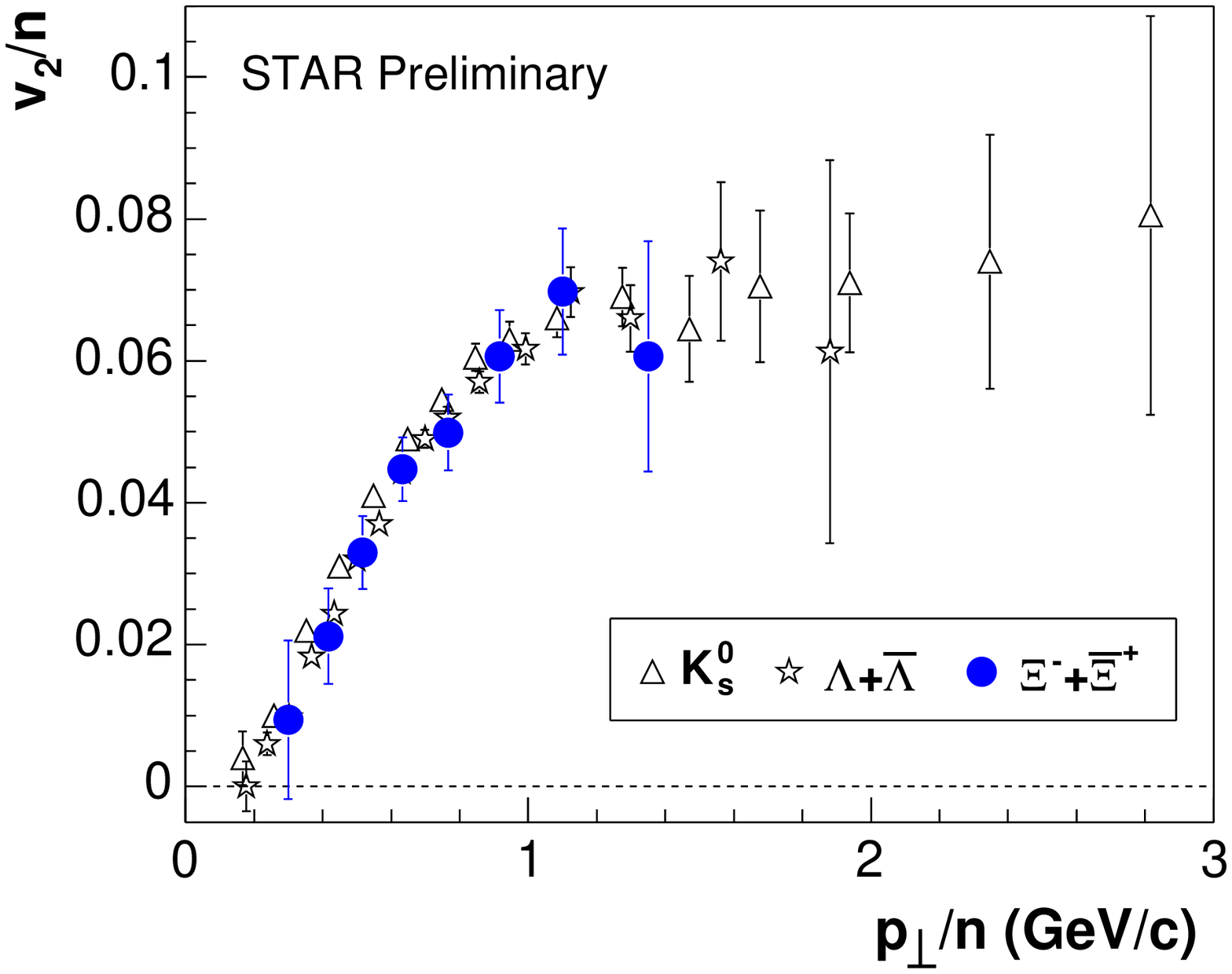,height=6cm}}
\vspace*{-10mm}
\caption{Elliptic flow $v_{2}$ of $K_{s}^{0}$, $\Lambda+\bar{\Lambda}$ and $\Xi^{-}+\bar{\Xi}^{+}$ normalized to the number of constituent quarks (n) as a function of $p_{T}/n$.}
\label{fig:v2nPtn}
\vspace*{-6mm}
\end{minipage}
\end{figure}

So this radial flow scenario suggests that for multi-strange baryons, a significant fraction (if not all) of the transverse flow has been developed probably in a partonic phase of the system so that multi-strange baryons should develop elliptic flow. Figure~\ref{fig:v2Pt} shows the measurement of the elliptic flow $v_{2}$ of $\Xi^{-}+\bar{\Xi}^{+}$ and $\Omega^{-}+\bar{\Omega}^{+}$ as a function of $p_{T}$ for the minimum bias data. $v_{2}$ of $K_{s}^{0}$ and $\Lambda+\bar{\Lambda}$ previously measured \cite{Soer04} are also represented for comparison.
\vspace*{-6mm}

First we observe that $v_{2}$ of multi-strange baryons is different from zero and seems to follow the same behaviour as $\Lambda$ $v_{2}$. That means same shape (saturation at a $p_{T}\sim3GeV/c$) and same amplitude (saturation at $v_{2}\sim20\%$). In the low $p_{T}$ region, $\Xi$ $v_{2}$ is in agreement with hydrodynamic model calculations (colored zone) which predict its mass ordering in this $p_{T}$ region. However, for a $p_{T}>2GeV/c$, $v_{2}$ deviates from Hydrodynamic model prediction and shows different behaviour for $K_{s}^{0}$ which saturates at a $p_{T}=2GeV/c$, at a value around $14\%$ compared to the strange baryons $v_{2}$.
It confirms a previously established baryon to meson dependence of the elliptic flow parameter from a particle mass dependence in the intermediate $p_{T}$ region \cite{Soer04}. This particle type dependence is well and ``simply'' explained by quark coalescence or recombination models \cite{Moln03,Frie03} in which hadrons are dominantly produced by the coalescence of constituent quarks from a partonic system supporting the idea of a collectivity between partons. These models predict a universal scaling of transverse momentum $p_{T}$ and elliptic flow to the number of constituent quarks ($n$). Previously, such scaling has been demonstrated for the mesons $K^{0}_{s}$ and the baryons $\Lambda$ at intermediate $p_{T}$ \cite{Soer04}. Figure~\ref{fig:v2nPtn} shows the superposition of the scaled elliptic flows $v_2/n=f(p_{T})/n$ for $K^{0}_{s}$, $\Lambda$ as well as for $\Xi^{-}+\overline{\Xi}^{+}$, supporting that the flow of $s$ quarks is close to that of $u$ and $d$ quarks within error bars.

\section{CONCLUSION}
We have presented the evolution of freeze-out parameters with Au+Au collision centrality. The increase of the strange quark phase space saturation factor, $\gamma_{s}$, up to 1 for the most central collision suggests that strangeness equilibration is achieved at top RHIC energy. $T_{ch}$, common for all particles, appears to be independent of the system size. The kinetic FO parameters obtained from blastwave fit to $\pi$, $K$, $p$ spectra suggest that they are taking part to a same collective flow behaviour with an increasing duration time between chemical and thermal FO with centrality essentially due to hadron rescatterings. For multi-strange baryons, the collective behaviour seems to be quite different. Since $T_{ch}$ and $T_{fo}$ are close to each other and show no dependence with centrality, it indicates that multi-strange baryons take less part in the evolution dynamics in the hadronic phase and should have decoupled much earlier in the collision than ($\pi$, $K$, $p$), carrying with them an important partonic flow contribution. This idea is emphasized by the measurement of their elliptic flow whose scaling by the constituent quarks in the intermediate $p_{T}$ region is well described by coalescence and recombination models.

\end{document}